\newcommand{\bra}[1]{\langle #1 \vert}
\newcommand{\ket}[1]{\vert #1 \rangle}

\documentclass[12pt]{iopart}

\expandafter\let\csname equation*\endcsname\relax
\expandafter\let\csname endequation*\endcsname\relax

\usepackage{dsfont}
\usepackage{blindtext}
\usepackage{graphicx}
\usepackage{epstopdf}
\usepackage{booktabs}
\usepackage{ulem}
\usepackage{tikz}
\usepackage{multirow}
\usepackage{emptypage}
\usepackage{physics}
\usepackage{gensymb}
\usepackage{appendix}
\usepackage{xpatch}
\newcommand{\lorena}[1]{{\color[rgb]{0.,0.,0.}{#1}}}
\newcommand{\yc}[1]{{\color[rgb]{0.,0.,0.}{#1}}}
\usepackage{hyperref}
\usepackage{xcolor}

\begin{document}

\title[Exploring weak value arguments and Bargmann invariants through the Majorana representation]{Exploring weak value arguments and Bargmann invariants in $N$-level quantum systems through the Majorana symmetric representation}

\author{Lorena Ballesteros Ferraz$^{1,2}$, Dominique Lambert$^3$ and Yves Caudano$^1$}

\address{$^1$Research Unit Lasers and Spectroscopies (UR-LLS), naXys \& NISM, University of Namur, Rue de
Bruxelles 61, B-5000 Namur, Belgium}
\address{$^2$ Laboratoire de Physique Théorique et Modélisation, CNRS Unité 8089,
CY Cergy Paris Université, 95302 Cergy-Pontoise cedex, France}
\address{$^3$ ESPHIN \& naXys, University of Namur, Rue de Bruxelles 61, B-5000 Namur, Belgium}
\ead{lorena.ballesteros-ferraz@cyu.fr, yves.caudano@unamur.be}

\begin{abstract}
This work examines the argument of weak values for general observables and develops a geometric description on the Bloch sphere. We apply the Majorana symmetric representation to reach this goal. The weak value of a general observable is proportional to the weak value of an effective projector: it is constructed from the application of the observable over the initial state, after normalization by a constant of proportionality that is real. The argument of the weak value of a projector on a pure state of an $N$-level system corresponds to a symplectic area in the complex projective space $(\text{CP}^{N-1})$. This symplectic area cannot be visualized directly but it can be represented geometrically with a sum of $N-1$ solid angles on the Bloch sphere using the Majorana stellar representation. By combining these two ideas, we show that the argument of the weak value of any observable (i.e., not just projectors) can be described with the Majorana representation, as the sum of $N-1$ solid angles on the Bloch sphere. These two approaches provide two geometrical descriptions, a first one in the complex projective space $\text{CP}^{N-1}$ and a second one on the Bloch sphere, after mapping the problem from the original $N$-dimensional quantum state space $(\text{CP}^{N-1})$ to a multi-qubit description in three-dimensional space by making use of the Majorana representation. These results can also be applied to the argument of the third-order Bargmann invariant, the most fundamental order as the argument of any higher order invariant can be expressed as a sum of the argument of third-order Bargmann invariants, as well as to the argument of the Kirkwood-Dirac quasi-probability distribution. Finally, we focus on the argument of the weak value of a general spin-1 operator when its modulus diverges towards infinity. This divergence amplifies signals with great usefulness in experiments and appears connected to the qubit entanglement in the Majorana representation. 
\end{abstract}

%
%
%
%
%

\maketitle

\section{Introduction}
Weak values have garnered significant attention for their key role in many areas of quantum physics, providing not only insights into questions relevant to quantum foundations but practical advantages in fundamental and applied experiments as well \cite{starling2010precision, pusey2014anomalous, dressel2012weak}.\\
Weak values arise notably when performing weakly a measurement of an observable through a unitary operator, followed by post-selection (executing a projective measurement and filtering the final state). Aharanov, Albert, and Vaidman defined this quantity in the context of the von Neumann scheme \cite{aharonov1988result, mello2014neumann}. In this protocol, the global system is composed by a measuring device (or ancilla) and a system of interest. The system and the ancilla interact through a unitary operator, $\hat{U}=\exp{-ig\hat{A}\otimes\hat{P}}$, where $\hat{A}$, acting on the system space, is the operator to be measured and $\hat{P}$, representing the momentum operator, acts on the measuring device space. After this interaction, the wave function of the measuring probe becomes a linear combination of wave functions that are shifted by quantities proportional to each of the eigenvalues of the observable \cite{svensson2013pedagogical, jozsa2007complex}. When the interaction strength, $g$, is small, the measurement is weak, and, in the absence of post-selection, the average shift in the ancilla's wave function is proportional to the expectation value of the observable $\hat{A}$. However, when post-selection is executed on the system after the weak interaction, all shifted wave functions are projected on a common state and thus interfere. As a result, the measuring device's wave function is typically shifted in position by a quantity proportional to the real part of the weak value, 
\begin{equation}
\label{eq:weak_value}
A_w=\frac{\bra{\psi_f}\hat{A}\ket{\psi_i}}{\bra{\psi_f}\ket{\psi_i}},
\end{equation}
where $\ket{\psi_i}$ and $\ket{\psi_f}$ are the pre- and post-selected states. The ancilla's wave function is simultaneously shifted in momentum by a quantity that is proportional to the imaginary part of the weak value \cite{svensson2013pedagogical, jozsa2007complex}.\\
Several schemes extended the weak measurement protocol beyond the specific configuration in which they were defined \cite{aharonov1988result}. The pointer can be any observable in continuous or discrete space. Furthermore, any observable can play the role of the probe, but the resulting shift is a linear combination of the real and imaginary parts \cite{svensson2013pedagogical}. Additionally, weak values can arise in more general schemes than weak measurements, for example, without a probe \cite{wiseman2002weak, ogawa2020operational},  in strong measurements \cite{de2022role, cormann2016revealing}, as a result of interference phenomena \cite{dressel2015weak}, or as classical background fields \cite{dressel_classical_2014}.\\  
Applications of weak measurements abound in different areas. Weak values are unbounded numbers; they enhance tiny signals \cite{dixon2009ultrasensitive, alves2015weak, xu2020approaching} and hence are a useful tool for sensing \lorena{\cite{zilberberg2011charge, luo2017precision, jordan2019gravitational, tang2025high}}, including in dissipative quantum systems \cite{ferraz2024relevance}. As complex numbers, they can be used in tomography \cite{hofmann2010complete, lundeen2012procedure, kim2018direct, zhu2017direct, lundeen2011direct}, for measuring wave functions \lorena{\cite{lundeen2011direct, gao2025wave}} and for measuring the expectation value of non-Hermitian operators \cite{pati2015measuring}. They also present a great potential in quantum computing \cite{lund2011efficient}. \\
Scientists usually study weak values in terms of their real and imaginary parts \cite{jozsa2007complex} because these quantities determine the observable meter shifts in typical experiments. Nonetheless, to provide a geometrical interpretation of these quantities, which has taken much attention over the last few years \cite{tamate2009geometrical, kedem2010modular, cormann2017geometric, ho2018various, samlan2017geometric, pal2019experimental, cho2019emergence, ferraz2023revisiting}, it is essential to investigate the argument. Anomalous weak values (complex values or values outside of the range of the possible expectation values of the observable) are proofs of contextuality, a ressource now viewed as essential to confer an advantage to quantum computing \cite{pusey2014anomalous, kunjwal2019anomalous} over classical computing. The argument of the weak value can thus help us to understand the meaning of weak values, based on a non-controversial, and hopefully intuitive, geometric picture. From a practical point of view, the argument provides the direction in which the coherent state of a Gaussian meter is shifted in phase space after post-selection in a weak measurement. Moreover, the real part of weak values is linked to the optimal conditional estimate of the observable, while the imaginary part is related to the inaccuracy of the estimate \cite{hall2004prior, dressel2015weak, dressel2012significance, hofmann2011uncertainty}. In consequence, the argument appears connected to a ratio of the estimate and its contribution to the error. In general, studying the geometric phase arising from weak values can benefit the study of interferometric phenomena with post-selection \lorena{\cite{starling2010precision, dressel2015weak, denkmayr2018weak, huang2025enhancing, masiello2025simultaneous}}. In the context of quantum thermodynamics, the trajectory-dependent stochastic entropy production is directly influenced by the argument of the weak value, under specific conditions. The argument plays a very significant role there, as a non-real trajectory-dependent stochastic entropy serves as a witness to non-classicality \cite{upadhyaya2024non}. Geometrically analyzing the argument of weak values becomes increasingly challenging as the system size grows, making visualization difficult even for relatively simple two-qubit systems. While prior work addressed this for projectors \cite{cormann2017geometric}, general discrete observables and their phase geometry on the Bloch sphere remained inaccessible until now. This work overcomes that limitation, providing a framework to study the argument of weak values in a parameter-dependent manner on the Bloch sphere. By analyzing the structural properties of weak value arguments and assessing their robustness—specifically, how much they fluctuate when parameters are adjusted—we can uncover new opportunities for applications in quantum information and foundational studies.\\
The argument of the weak value of any two-level projector is associated to a geometric phase that is proportional to the solid angle on the Bloch sphere of the spherical triangle spanned by the pre-selected state, the projector state and the post-selected state \cite{cormann2016revealing}. For $N$-level systems, the argument of the weak value of a projector represents a geometric phase associated to the symplectic area of the geodesic triangle spanned by the pre-selected state, the projector state, and the post-selected state in $\text{CP}^{N-1}$ \cite{ballesteros2022geometrical}. However, the symplectic area in $\text{CP}^{N-1}$ ($N\neq2$) does not correspond to a solid angle in a larger dimensional space; thus, its visual representation is not straightforward. The complex projective space $\text{CP}^{N-1}$ is the natural mathematical setting for describing the projectors corresponding to pure quantum states in an $N$-dimensional Hilbert space. It captures all the physical parameters of pure states, excluding the unobservable global phase and the arbitrary amplitude of vectors in Hilbert space. For a two-level quantum system, $\text{CP}^{1}$ can be visualized as the surface of the Bloch sphere (the usual sphere in 3 dimensions, noted  $S^2$), providing an intuitive geometric representation of the state space. As it was recently shown that the argument of the weak value of any observable is equivalent to the argument of the weak value of an effective projector (an extra phase is involved, with $0$ or $\pi$ as sole possible values) \cite{ballesteros2022geometrical}, we can apply the geometric description previously developed for projectors to weak values of general observables. 
\yc{Using the generators of $\textrm{SU}(N)$, it is possible to represent pure quantum states of $N$-level systems as real vectors defining points on the surface of the $S^{N^2-2}$ sphere in $N^2-1$ dimensions. However, states only correspond to a $2N-2$ dimensional subspace of that hypersphere, so that a majority of the sphere surface does not correspond to quantum states when $N>2$. While the argument of the weak value in a qubit system corresponds to a solid angle on the Bloch sphere, this property does not generalize in higher dimensions, as the argument of the weak value does not appear associated to a (generalized) solid angle intercepting the $S^{N^2-2}$ sphere. We thus seek an alternative representation.}\\
Cormann et al. showed that the argument of the weak value of $N$-level projectors can be expressed as the sum of $N-1$ solid angles on the Bloch sphere. For this, they applied Majorana representation to the three states (initial state, projector state, and post-selected state)\cite{cormann2017geometric}. This work focused on projectors, which are not representative of the majority of measured observables, notably spin or energy. We stress that our work has a much larger application range, as it applies to all discrete observables and, beyond the context of weak values, to Bargmann invariants, complex-valued quantity derived from the overlaps of quantum states that encapsulates geometric and topological properties of quantum systems in Hilbert space, as well as the Kirkwood-Dirac quasi-probability distribution, which has seen renewed and considerable interest in the past  15 years \lorena{\cite{das2023saturating, arvidsson-shukur_properties_2024, wagner2024quantum}}.\\
Majorana introduced in the 1930s a mathematical procedure to represent systems larger than qubits on the Bloch sphere. $N-1$ stars on the Bloch sphere represent an $N$-level system \cite{majorana1932atomi}. The simple intuition behind this representation stems from the isomorphism existing between the Hilbert space of an $N$-dimensional quantum system and the Hilbert subspace of the symmetric states of $N-1$ qubits. As an illustration, let us consider two spin-$1/2$ qubits. By the laws of addition of angular momentum, the total angular momentum can either be $J=0$, corresponding to a one-dimensional anti-symmetric state (the $\ket{\Psi^-}=(\ket{\uparrow\downarrow}-\ket{\downarrow\uparrow})/\sqrt{2}$ Bell state), or $J=1$, corresponding to a three-dimensional symmetric subspace (with basis eigenstates $\ket{\uparrow\uparrow}\ (J_Z=1)$, $\ket{\Psi^+}=(\ket{\uparrow\downarrow}+\ket{\downarrow\uparrow})/\sqrt{2}\ (J_Z=0)$, and $\ket{\downarrow\downarrow}\ (J_Z=-1)$, all symmetric under the permutation of the two qubit states). In practice, this representation maps the states of the $N$-level system in consideration to the $J=(N-1)/2$ angular momentum subspace of $N-1$ qubits, which is also of dimension $N$. Consequently, the different qubit states associated to the equivalent, fully-symmetric, angular momentum representation of the $N$-level quantum state can be drawn on the Bloch sphere, as a Majorana star constellation. The Majorana representation is a powerful tool to get a geometrical insight and to perform calculations \cite{hannay1998berry, bliokh2019geometric, markham2011entanglement}. Several studies, from purely theoretical to quantum computing, made use of this representation \cite{zimba2006anticoherent, galindo2022entangling, akhilesh2019spin}.\\
Visualizing the geometry of the studied system is essential, specially, from a dynamical point of view, perhaps, in terms of the final or initial state. Nonetheless, visualizing a symplectic area in the complex projective space $\text{CP}^{N-1}$ is not an intuitive task, as it is not the usual Riemannian area. To tackle this problem, in this paper, we show that, by applying the Majorana representation to the three states involved in the weak value of any $N$-level observable (pre-selected state, effective projector state \cite{ballesteros2022geometrical}, and post-selected state), the geometry of the full system is brought to the Bloch sphere. The argument of the weak value is the sum of $N-1$ solid angles on the Bloch sphere. The argument of the projector weak value is equivalent to the argument of the Bargmann invariant, $\arg{\Delta_B}$ associated to three states (invariant under gauge transformation and re-parametrization) that can be expressed as
\begin{equation}
\arg{\Delta_B}=\arg{\text{Tr}\left[\hat{\Pi}_1\hat{\Pi}_2\hat{\Pi}_3\right]},
\end{equation}\\
where each $\hat{\Pi}_i$ represents a projector \cite{ballesteros2022geometrical}. \lorena{In this article, we focus our study on}\yc{ weak values of} \lorena{discrete observables}\yc{ with pre- and post-selection}\lorena{ in pure states.} The geometric interpretation is thus also appropriate in this context. Bargmann invariants are directly linked to the Kirkwood-Dirac quasi-probability distribution that defines a non-classical state by taking negative or complex values \yc{\cite{arvidsson-shukur_properties_2024, kirkwood1933quantum}}. Third-order Bargmann invariants, $\text{Tr}\left(\hat{\Pi}_1\hat{\Pi}_2\hat{\Pi}_3\right)$, are especially interesting, as the argument of any $N$-order Bargmann invariant, $\text{Tr}\left(\hat{\Pi}_1\hat{\Pi}_2\hdots\hat{\Pi}_N\right)$ can be expressed as the sum of $N-2$ arguments of third-order Bargmann invariants \cite{rabei1999bargmann}. The weak value is equal to a third-order Bargmann invariant divided by the projection probability of the pre- and post-selected states. \\
This paper is structured as follows. \lorena{In section~\ref{section:Weak_values_of_N-level_observables_in_terms_of_Majorana_stars}}, we present the geometric interpretation of weak values of $N$-level general observables, by applying the Majorana representation. \lorena{In section~\ref{section:Majorana_three_level_system}}, these calculations are applied to the specific case of $3$-level systems. After this\lorena{, in section~\ref{section:spin_1}}, we present the relevant example of spin-$1$ systems: we study the argument of the weak value of a spin-1 operator when the modulus of the weak value presents a divergence, a typical situation of an amplification effect appearing in a weak measurement with nearly orthogonal pre- and post-selected states. \lorena{Finally, we present our conclusions (section~\ref{section:conclusions}).}\\

\section{Weak values of $N$-level observables in terms of Majorana stars}\label{section:Weak_values_of_N-level_observables_in_terms_of_Majorana_stars}
The initial state, the observable and the post-selected state constitute the required components of a weak value. Varying any of these parts can completely modify the quantity. In this section, we provide the theoretical framework to apply the Majorana representation to the different components of the weak value of an $N$-level observable.\\
The weak value of any discrete observable is proportional to the weak value of a very specific projector with a constant of proportionality that is real \cite{ballesteros2022geometrical}, 
\begin{equation}
\label{eq:definition_weak_value_i_p}
A_w=\frac{\bra{\psi_f}\hat{A}\ket{\psi_i}}{\bra{\psi_f}\ket{\psi_i}}=\frac{\bra{\psi_i}\hat{A}^2\ket{\psi_i}}{\bra{\psi_i}\hat{A}\ket{\psi_i}}\frac{\bra{\psi_f}\hat{\Pi}_{i'}\ket{\psi_i}}{\bra{\psi_f}\ket{\psi_i}},
\end{equation}
\\
where $\hat{\Pi}_{i'}=\ket{\psi_{i'}}\bra{\psi_{i'}}$, with,
\begin{equation}
\label{eq:equation_psi_i_prime}
\ket{\psi_{i'}}=\frac{1}{\sqrt{\bra{\psi_{i}}\hat{A}^2\ket{\psi_i}}}\hat{A}\ket{\psi_{i}}, 
\end{equation}
Eq.~(\ref{eq:definition_weak_value_i_p}) does not present any issue of definition. When $\langle\hat{A}^2\rangle_{\psi_{i}}=0$, the weak value $A_w$ is equal to $0$.  If the expectation value of $\hat{A}$ in the initial states is equal to $0$, then the weak value of the effective projector is calculated through a limit, using a small parameter $\epsilon$ that tends to $0$ (for more details, see \cite{ballesteros2022geometrical}). As a result, any weak value can be fully described by the three states $\ket{\psi_i}$, $\ket{\psi_{i^\prime}}$, and $\ket{\psi_f}$.\\
Since the weak value remains invariant under unitary transformations, two unitary operators are applied to transform two of the three states defining the weak value into separable states in the Majorana representation. \lorena{Any two states can always be mapped to separable states in the Majorana representation \cite{tamate2011bloch}, which correspond to degenerate stars—i.e., a coherent state. Coherent states are characterized by the maximal degeneration of the stars, with all stars occupying the same position on the sphere.} Indeed, there always exist two coherent states $\ket{\Psi}$ and $\ket{\Phi}$ in the Majorana representation whose scalar product equals the scalar product of the two chosen states $\ket{\psi}$ and $\ket{\phi}$ of the $N$-level system: $\langle \phi \vert \psi \rangle=\langle \Phi \vert \Psi \rangle=\langle \beta_\Phi \vert \alpha_\Psi \rangle^{N-1}$, with $\ket{\alpha_\Psi}$ and $\ket{\beta_\Phi}$ corresponding to the degenerate qubit states associated with $\ket{\Psi}$ and $\ket{\Phi}$, respectively. The conservation of the scalar product implies the existence of a unitary mapping between the states \cite{tamate2011bloch}. For a more detailed explanation of the application of the Majorana symmetric representation, see~\ref{section:appendix_2}. \\
For simplicity, the pre-selected state is mapped to,
\begin{equation}\label{eq:projection-symmetry-unitary}
\ket{\psi_i}\rightarrow\ket{\Psi'_i}=\underbrace{\ket{\uparrow}\ldots\ket{\uparrow}}_{N-1},
\end{equation}
via applying an appropriate unitary operator $\hat{U}^{\left(1\right)}$. In the original Hilbert space, $\hat{U}^{(1)}$ maps $\ket{\psi_i}\rightarrow\ket{\psi_i^\prime}=\ket{0}$. Throughout this work, we will use $\left\{\ket{\uparrow},\ket{\downarrow}\right\}$ for the qubit bases, and $\left\{\ket{0}, \ket{1},\ldots,\ket{N-1}\right\}$ for the original Hilbert space basis\footnote{With $N=1$, the correspondance is $\ket{\uparrow}=\ket{0}$ and $\ket{\downarrow}=\ket{1}$.}.
The general form of the unitary operator to take a state $\ket{\psi_1}$ to another state $\ket{\psi_2}$ is,
\begin{equation}
\label{eq:projection_symmetry_unitary}
\hat{U}=e^{-j\text{arg}\bra{\psi_2}\ket{\psi_1}}\left(\hat{I}-2\ket{\Delta}\bra{\Delta}\right), 
\end{equation}
where,
\begin{equation}
\label{eq:state_of_projection_symmetry_unitary}
\ket{\Delta}=\frac{e^{-j\text{arg}\bra{\psi_2}\ket{\psi_1}}\ket{\psi_1}-\ket{\psi_2}}{\sqrt{2\left(1-|\bra{\psi_2}\ket{\psi_1}|\right)}}.
\end{equation}
Here, we set thus $\ket{\psi_1}=\ket{\psi_i}$ and $\ket{\psi_2}=\ket{0}$. The other components are also affected by the unitary transformation, $\hat{A}\rightarrow\hat{U}^{\left(1\right)}{\hat{A}\hat{U}^{\left(1\right)}}^{\dagger}=\hat{A'}$, and $\hat{U}^{\left(1\right)}\ket{\psi_f}=\ket{\psi^\prime_f}$. \\
A second unitary operator $\hat{U}^{\left(2\right)}$ that leaves the pre-selected state invariant, $\ket{\psi''_i}=\hat{U}^{\left(2\right)}\ket{\psi'_i}=\ket{\psi'_i}$, is applied to map the state \lorena{$\ket{\psi'_{i'}}$} to a second separable state, 
\begin{equation}
\label{eq:after_second_unitary_general}
\ket{\psi'_{i'}}\rightarrow\ket{\Psi''_{i'}}=e^{j \mathrm{arg}\langle \psi_i\vert\psi_{i^\prime}\rangle}\underbrace{\ket{\phi_{i'}}\ldots\ket{\phi_{i'}}}_{N-1},
\end{equation}
where the qubit state takes the form $\ket{\phi_{i^\prime}}=\cos \frac{\theta}{2} \ket{\uparrow}+e^{j \varphi}\sin\frac{\theta}{2} \ket{\downarrow}$. The phase factor in (\ref{eq:after_second_unitary_general}) is required to preserve the scalar product $\langle \psi_i\vert\psi_{i^\prime}\rangle$. If desired, an appropriate reparametrization of  the original Hilbert space can set a positive real value for the state overlap $\langle \psi_i\vert\psi_{i^\prime}\rangle=\vert\langle \psi_i\vert\psi_{i^\prime}\rangle\vert$, without loss of generality\footnote{With the definition (\ref{eq:equation_psi_i_prime}) of $\ket{\psi_{i^\prime}}$, the overlap $\langle \psi_i\vert\psi_{i^\prime}\rangle=\langle\hat{A}\rangle_{\psi_i}/\sqrt{\langle\hat{A}^2\rangle_{\psi_i}}$ is a real number when $\hat{A}$ is a Hermitian observable. Nevertheless, we would like to keep the procedure general at this stage: thus we will not make any assumption on the specific form taken by the state $\ket{\psi_{i^\prime}}$.}. In the original Hilbert space, $\hat{U}^{(2)}$ maps $\ket{\psi_{i^\prime}^\prime}\rightarrow\ket{\psi_{i^\prime}^{\prime\prime}}=e^{j \mathrm{arg}\langle \psi_i\vert\psi_{i^\prime}\rangle}\sum_{k=0}^{N-1} d_k \ket{k}$, with coefficients $d_k$ to be determined from the expansion of the state $\ket{\Psi''_{i'}}$ in the symmetric Dicke basis, using the Majorana isomorphism
\begin{eqnarray}
\sum_{k=0}^{N-1} d_k \ket{k}\cong\left(\cos\frac{\theta}{2}\ket{\uparrow}+e^{j \varphi} \sin\frac{\theta}{2}\ket{\downarrow}\right)^{\otimes (N-1)}\nonumber\\
= \sum_{k=0}^{N-1} \sqrt{C_k^{N-1}}\left(\cos\frac{\theta}{2}\right)^{N-1-k}\left(e^{j \varphi} \sin\frac{\theta}{2}\right)^k\ket{\mathcal{D}^{N-1}_{k}}
\end{eqnarray}
with the Dicke states defined by (see also~\ref{section:appendix_2}) 
\begin{equation}\label{eq.Dicke}
\ket{\mathcal{D}^{N-1}_{k}}=\frac{1}{\sqrt{C_{k}^{N-1}}}\sum_{P} \hat{P}\left[\ket{\uparrow}^{\otimes (N-1-k)}\ket{\downarrow}^{\otimes k}\right],
\end{equation}
where $C_k^{N-1}=\frac{\left(N-1\right)!}{k!\left(N-1-k\right)!}$ represent the binomial coefficients, where the sum runs through all the permutations $\hat{P}$ of the
qubit states, and where the phase $e^{j \mathrm{arg}\langle \psi_i\vert\psi_{i^\prime}\rangle}$ was factored out of the coefficients $d_k$ for convenience in the definition of $\ket{\psi_{i^\prime}^{\prime\prime}}$. By comparing the state overlap in the Majorana representation $\langle \Psi_i^{\prime\prime}\vert\Psi_{i^\prime}^{\prime\prime}\rangle=e^{j \mathrm{arg}\langle \psi_i\vert\psi_{i^\prime}\rangle} \langle \uparrow \vert\phi_{i^\prime}\rangle^{N-1}=e^{j \mathrm{arg}\langle \psi_i\vert\psi_{i^\prime}\rangle} (\cos\frac{\theta}{2})^{N-1}$ with its expression in the original Hilbert space $\langle \psi_i\vert\psi_{i^\prime}\rangle=\langle \psi_i^{\prime\prime}\vert\psi_{i^\prime}^{\prime\prime}\rangle=\langle 0\vert\psi_{i^\prime}^{\prime\prime}\rangle=e^{j \mathrm{arg}\langle \psi_i\vert\psi_{i^\prime}\rangle} d_0$, we find $d_0=\vert\langle \psi_i\vert\psi_{i^\prime}\rangle\vert=(\cos\frac{\theta}{2})^{N-1}$. Therefore, we determine the angle $\theta=2\arccos\sqrt[N-1]{\vert\langle \psi_i\vert\psi_{i^\prime}\rangle\vert}$. In practice, we can set $\varphi=0$ by an adequate rotation of the Bloch sphere around its vertical axis, effectively mapping $\ket{\psi_{i^\prime}^{\prime\prime}}$ on the 0 longitude on the Bloch sphere. As a result the state $\ket{\psi_{i^\prime}^{\prime\prime}}$ is completely determined from the overlap:
\begin{eqnarray}\label{eq:psi_iprime_second_expression1}
d_k&=&\sqrt{C_k^{N-1}}\left(\cos\frac{\theta}{2}\right)^{N-1-k}\left(\sin\frac{\theta}{2}\right)^k\\
   &=&\sqrt{C_k^{N-1}}\vert\langle \psi_i\vert\psi_{i^\prime}\rangle\vert^{\frac{N-1-k}{N-1}} \sqrt{1-\vert\langle \psi_i\vert\psi_{i^\prime}\rangle\vert^{\frac{2}{N-1}}}^k.\label{eq:psi_iprime_second_expression2}
\end{eqnarray}
With the knowledge of $\ket{\psi_{i^\prime}^{\prime\prime}}$, we can construct the unitary $\hat{U}^{(2)}$ by following a procedure similar to the one outlined in equations (\ref{eq:projection_symmetry_unitary}) and (\ref{eq:state_of_projection_symmetry_unitary}):
\begin{equation}
\hat{U}^{(2)}=\ket{0}\bra{0}+e^{-j\text{arg}\bra{\psi_2}\ket{\psi_1}}\left(\hat{I}-\ket{0}\bra{0}-2\ket{\Delta}\bra{\Delta}\right),
\end{equation}
where $\ket{\Delta}$ is given by (\ref{eq:state_of_projection_symmetry_unitary}) with the states $\ket{\psi_1}$ and $\ket{\psi_2}$ corresponding now to the components of $\ket{\psi_{i^\prime}^{\prime}}$ and $\ket{\psi_{i^\prime}^{\prime\prime}}$ that are orthogonal to $\ket{0}$, i. e.
\begin{equation}
\ket{\psi_1}= \frac{\ket{\psi_{i^\prime}^{\prime}}-\langle 0 \vert \psi_{i^\prime}^{\prime}\rangle \ket{0}}{\sqrt{1-\vert\langle 0 \vert\psi_{i^\prime}^{\prime}\rangle\vert^2}}, \quad
\ket{\psi_2}= \frac{\ket{\psi_{i^\prime}^{\prime\prime}}-\langle 0 \vert \psi_{i^\prime}^{\prime\prime}\rangle \ket{0}}{\sqrt{1-\vert\langle 0 \vert\psi_{i^\prime}^{\prime\prime}\rangle\vert^2}}
\end{equation}
with $\ket{\psi_{i^\prime}^{\prime\prime}}$ determined from (\ref{eq:psi_iprime_second_expression2}). It is easy to check that $\hat{U}^{(2)}$ leaves the state $\ket{\psi_i^{\prime}}=\ket{0}$ invariant.\\
The unitary operator $\hat{U}^{(2)}$ should also be applied to the post-selected state, $\hat{U}^{\left(2\right)}\ket{\psi'_{f}}=\ket{\psi''_{f}}$. After both unitary operators, the post-selected state becomes a general $N$-level state, 
\begin{equation}
\ket{\Psi''_{f}}=\frac{e^{j \mathrm{arg}\langle 0 \vert \psi_{f}^{\prime\prime}\rangle}}{\sqrt{M}}\sum_{P}\hat{P}\left[\ket{\phi_{f}^{\left(1\right)}}...\ket{\phi_{f}^{\left(N-1\right)}}\right], 
\end{equation}
where the sum runs through all the permutations $\hat{P}$ of the qubit states, and $M$ is a normalization constant. Writing the state as $\ket{\psi''_{f}}=e^{j \mathrm{arg}\langle 0 \vert \psi_{f}^{\prime\prime}\rangle}\sum_{i=0}^{N-1}c_i\ket{i}$ and solving the Majorana polynomial, Eq.~(\ref{eq:general_polynomial}), one can express the state in the Majorana symmetric representation \cite{bloch1945atoms, devi2012majorana}, \\
\begin{equation}
\label{eq:general_polynomial}
P\left(z\right)=\sum_{k=0}^{N-1}\left(-1\right)^{k}\sqrt{C_{k}^{N-1}}c_{k}z^{N-1-k}, 
\end{equation}
where $C_k^{N-1}=\frac{\left(N-1\right)!}{k!\left(N-1-k\right)!}$ represent the binomial coefficients, and $c_k$ are the coefficients of the state $\ket{\psi''_{f}}$. The polar, $\theta_k$, and azimuthal, $\phi_k$, angles on the Bloch sphere depend respectively on the modulus and the phase of the roots $z_k$ of the polynomial expressed in Eq.~(\ref{eq:general_polynomial}),
\begin{equation}
\label{eq:angles_roots_polynomial}
z_k=e^{j\phi_k}\tan{\frac{\theta_k}{2}},
\end{equation}\\
where $0\leq\phi_k\leq 2\pi$, and $0\leq\theta_k\leq\pi$. From a geometric perspective, $z_k$ represents the stereographic projection in the complex plane of the $k^{th}$ qubit state contributing to the Majorana representation \yc{(we considered a projection from the south pole)}. When equaled to zero, the Majorana polynomial expresses the orthogonality between the state $\sum_{i=0}^{N-1} c_i \ket{i}$ and the coherent state $\left(\cos\frac{\pi-\theta_k}{2}\ket{\uparrow}+\sin\frac{\pi-\theta_k}{2}\ e^{j \phi_k + j \pi}\ket{\downarrow}\right)^{\otimes (N-1)}$ orthogonal to the qubit described by $z_k$, indicating that the inverse stereographic projection of $z_k$ must belong to the Majorana constellation.\\
The weak value is now calculated, following Eq.~(\ref{eq:definition_weak_value_i_p}), as, 
\begin{equation}
A_w=\frac{\bra{\psi''_f}\hat{A}''\ket{\psi''_i}}{\bra{\psi''_f}\ket{\psi''_i}}=\frac{\bra{\psi_i}\hat{A}^2\ket{\psi_i}}{\bra{\psi_i}\hat{A}\ket{\psi_i}}\Pi_{i',w}^{\left(1\right)}\Pi_{i',w}^{\left(2\right)}\hdots\Pi_{i',w}^{\left(N-1\right)},
\end{equation}
where each two-level system weak value is $\Pi_{i',w}^{\left(\lorena{v}\right)}=\frac{\bra{\phi_{f}^{\left(\lorena{v}\right)}}\ket{\phi_{i'}}\bra{\phi_{i'}}\ket{\phi_{i}}}{\bra{\phi_{f}^{\left(\lorena{v}\right)}}\ket{\phi_{i}}}$.\\ 
The modulus of the weak value is thus the product of $N-1$ moduli of weak values of qubit projectors,
\begin{eqnarray}
&&|A_w|=\frac{\bra{\psi_i}\hat{A}^2\ket{\psi_i}}{|\bra{\psi_i}\hat{A}\ket{\psi_i}|}|\Pi_{i',w}^{\left(1\right)}|\cdot|\Pi_{i',w}^{\left(2\right)}|...|\Pi_{i',w}^{\left(N-1\right)}|\\ \nonumber
&=&\sqrt{\bra{\psi_i}\hat{A}^2\ket{\psi_i}}\left|\frac{\bra{\phi_{f}^{\left(1\right)}}\ket{\phi_{i'}}}{\bra{\phi_{f}^{\left(1\right)}}\ket{\phi_{i}}}\right|\left|\frac{\bra{\phi_{f}^{\left(2\right)}}\ket{\phi_{i'}}}{\bra{\phi_{f}^{\left(2\right)}}\ket{\phi_{i}}}\right|\hdots\left|\frac{\bra{\phi_{f}^{\left(N-1\right)}}\ket{\phi_{i'}}}{\bra{\phi_{f}^{\left(N-1\right)}}\ket{\phi_{i}}}\right|
\end{eqnarray}
\\
The argument of the weak value of $\hat{A}$ is the sum of $N-1$ arguments of weak values of qubit projectors and the argument of the expected value of the operator $\langle\hat{A}\rangle_i$, which is either $0$ or $\pi$.\\
\yc{\begin{eqnarray}
\text{arg}A_w&=&\text{arg}\Pi_{i',w}^{\left(1\right)}+\text{arg}\Pi_{i',w}^{\left(2\right)}+\ldots+\text{arg}\Pi_{i',w}^{\left(N-1\right)}-\text{arg}\langle\hat{A}\rangle_i, \\ \nonumber
&=&-\frac{\Omega_{ii'_1f}}{2}-\frac{\Omega_{ii'_2f}}{2}-\ldots-\frac{\Omega_{ii'_{N-1}f}}{2}-\text{arg}\langle\hat{A}\rangle_i\\ \nonumber
&=&\sum_{v}\text{arg}\left(\bra{\phi_{i}}\ket{\phi_{f}^{\left(v\right)}}\bra{\phi_{f}^{\left(v\right)}}\ket{\phi_{i'}}\right)-\text{arg}\langle\hat{A}\rangle_i\lorena{,}
\end{eqnarray}}
\lorena{where each $\Omega_{abc}$ is the oriented solid angle subtended at the center of the Bloch sphere by the geodesic triangle whose vertices are $\vec{a}$, $\vec{b}$, and $\vec{c}$.} In this expression, each argument of a qubit projector weak value corresponds to a geometric phase, which is associated with the area of a solid angle on the Bloch sphere. Each solid angle is in turn defined by the spherical triangle formed by the vectors representing the pre-selected qubit state, the effective qubit state resulting from the application of the observable on the initial state, and the post-selected qubit state in the Majorana representation \lorena{\cite{cormann2017geometric}}.\\ 
Thus, the argument of the weak value of an observable in an $N$-level system represents a geometric phase that is associated to the symplectic area of the geodesic triangle spanned by the geodesics linking the three vectors representing the pre-selected state, the application of the observable over the pre-selected state and the post-selected state in the complex projective space $\text{CP}^{N-1}$ describing the geometry of the original quantum state space. This space is a Kähler manifold, which means that there are three compatible structures: the complex structure, the symplectic structure and the Riemannian structure. In $\text{CP}^{1}$, the symplectic area and the Riemannian one coincide. Hence, the argument of the weak value of an observable in two-level systems can be described in terms of solid angles. Using Majorana's description, we succeed to associate a symplectic area in the complex projective space $\text{CP}^{N-1}$ with $N-1$ solid angles on the Bloch sphere. However, the spherical triangles associated to the solid angles, $\Omega$, do not correspond to geodesic curves of $\text{CP}^{N-1}$ in the Majorana representation \yc{\cite{mittal2022geometric}}. In other words, while the symplectic area determining the argument of the weak value is defined as a contour integral along the actual geodesics of the quantum state manifold, the visualization of this symplectic area on the Bloch sphere involves geodesics of the sphere, which, perhaps surprisingly, do not describe the original geodesics of the quantum state manifold\yc{\footnote{\yc{For example, ref. \cite{mittal2022geometric} shows that for a three-level system, the geodesics between two coherent states correspond to a circle on the sphere, with the two coherent states located at opposite ends of a diameter. Thus, the two half circle arcs joining the stars are not great circles arcs in general. Our numerical computations show that, in general, the geodesics are not even circle arcs.}}}. This simplifies significantly the description, which enables visualizing the argument of the weak value and Bargmann invariants as function of dynamical parameters, giving a direct intuition of the studied system. In general, Majorana stars do not directly correspond to physical qubits because all the states of an $N$-qubit system generate a $2^N$ dimensional Hilbert space represented by $2^N -1$ Majorana stars \cite{bengtsson2017geometry}. Sometimes, the studied systems are actually composed of $k$ different qubits in a symmetric state. In this case, the Majorana representation would provide the actual geometry on the Bloch sphere of the qubit states defining the global symmetric state due to the bijection linking the fully symmetric quantum states of the qubits and their depiction as a Majorana constellation on the Bloch sphere, which results from the isomorphism between the Hilbert space of arbitrary $N$-level quantum systems and the $N$-dimensional subspace of $N-1$ qubits with total angular momentum  $J=(N-1)/2$.
\section{Majorana representation of weak values of observables in three-level systems}\label{section:Majorana_three_level_system}
\begin{figure*}
\centering 
\includegraphics[width=1.0\textwidth]{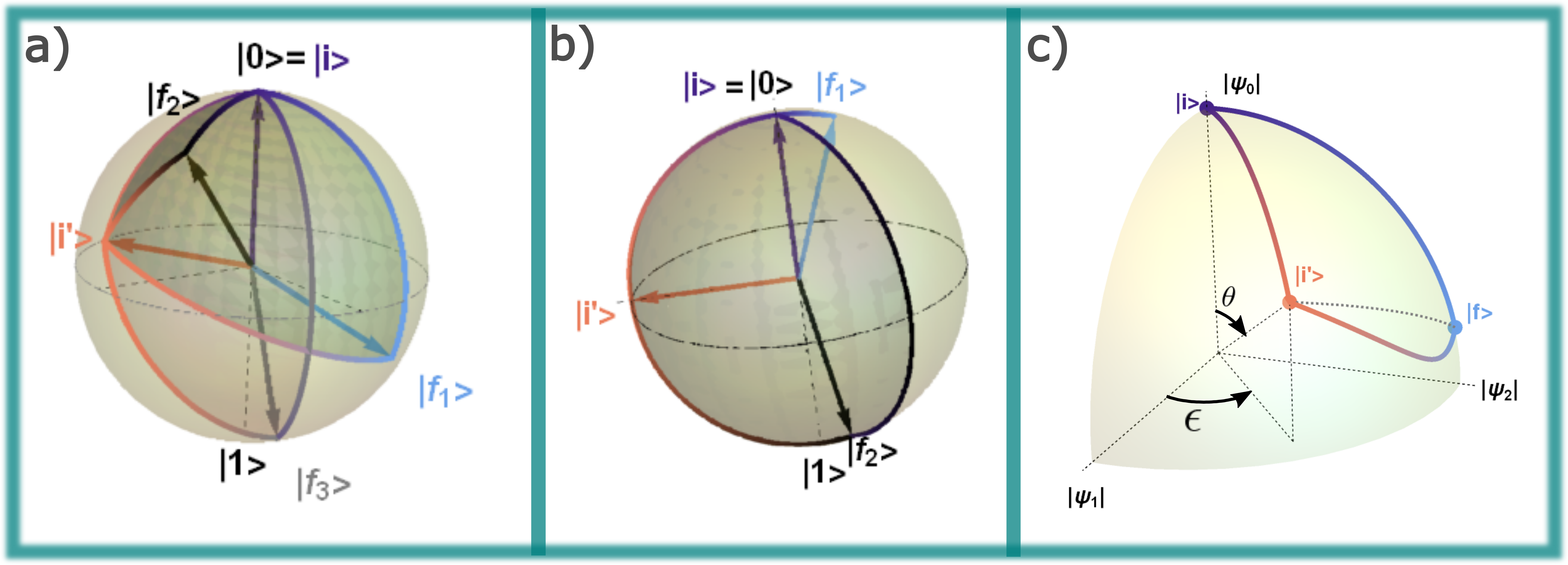}
  \caption{Representation of the argument of the weak value of the CNOT gate. The pre-selected state is $\ket{\psi_i}=\frac{1}{\sqrt{4}}\left(1, -j, 1, -j\right)^T$ and the post-selected state is $\ket{\psi_f}=\frac{1}{\sqrt{5}}\left(1, 0, -2, 0\right)^T$. a) Representation of the three solid angles involved in $\text{arg}\left(CNOT_w\right)$ on the Bloch sphere. b) Representation of two solid angles concerning the argument of the weak value of the $CNOT$ gate in the reduced approach (three-level system). c) Depiction of the geodesic triangle in the complex projective space $\text{CP}^2$ between the initial state, the application of the CNOT gate over the initial state and the post-selected state, using the spherical octant projection. By rotating, these figures can be easily visualized as a function of a varying parameter.   
	\label{fig:majorana_and_cp2}}
\end{figure*}

In this section, we focus on weak values of three-level general observables, defined as $\hat{A}=a_I\hat{I}+a_L\Vec{\alpha}\cdot\Vec{\hat{\lambda}}$, where $\Vec{\hat{\lambda}}$ are the Gell-Mann matrices (\ref{section:appendix_1}). We demonstrate the procedure using the Majorana representation, highlighting weak values and Bargmann invariants of three-level systems, which hold special significance. On the one hand, three-level systems feature several interesting observables, such as spin-1 operators, 3D Stokes parameter operators, or three-level projectors like those appearing in the three-box paradox \cite{bertlmann2008bloch, andreev20223d, ravon2007three, acin2002quantum}. On the other hand, as weak values are only dependent on three vectors, the description of a single weak value is intrinsically a three-level problem. Since weak values are invariant under unitary transformations, it is feasible to apply three unitary operators to transform the three $N$-level vectors into three states with only three components different from zero. In practice, this maps the vectors and their associated geodesic triangles to a three-dimensional subspace of the complex projective space $\text{CP}^{N-1}$, equivalent to $\text{CP}^2$. By applying this procedure, any weak value of systems larger than three dimensions can be converted to a three-level weak value, providing a representation of its argument of the weak value as two solid angles on the Bloch sphere. Consequently, we can always choose to represent the argument with $N-1$ or two solid angles.\\
Any projector of a pure three-level state can be written in terms of \lorena{a normalized vector whose components are the Gell-Mann matrices ordered as expressed in \ref{section:appendix_1}} and the identity as
\begin{equation}
\label{eq:projector_three-level_system}
\hat{\Pi}_{a}=\frac{1}{3}\left(\hat{I}+\sqrt{3}\Vec{a}\cdot\Vec{\hat{\lambda}}\right),
\end{equation}
\lorena{where the vector $\Vec{a}$ must satisfy the condition $\Vec{a}\star\Vec{a}=\Vec{a}$, where the symmetric star product is defined as $\left(\Vec{a}\star\Vec{b}\right)_k=\sqrt{3}d_{ijk}a_{i}b_{j}$, using the Einstein notation \cite{mallesh1997generalized, khanna1997geometric}}\yc{, with $d_{ijk}$ the symmetric structure constants of the group $\textrm{SU}(N)$ (which arise from the anti-commutation relations of the $N^2-1$ generators $\hat{\lambda}_i$)}.\\
The weak value of $\hat{A}$, $A_w$ defined as in Eq.~\ref{eq:weak_value}, is proportional to the weak value of the projector $\hat{\Pi}_{i'}$, where $\ket{\psi_{i'}}$ is defined in Eq.~(\ref{eq:equation_psi_i_prime}). Owing to this property, the weak values of general observables are directly linked to Bargmann invariants, as explained in the previous sections. The argument of the weak value is equal to the argument of a Bargmann invariant up to a phase of either $0$ or $\pi$.\\
With a description of weak values for general observables in terms of projectors, the Majorana representation can be applied to all three states. In this framework, the system is mapped from $\text{CP}^{2}$ to a representation on the Bloch sphere. The argument of the weak value of a projector is the sum of the arguments of two weak values in two-level systems. Each of these arguments is associated to a solid angle on the Bloch sphere.\\
Let us consider a general pre-selected state $\ket{\psi_{i}}$ in $\text{CP}^{2}$ (removing the global phase), 
\begin{equation}
\label{eq:expression_psi_i_in_terms_of_vectors}
\ket{\psi_{i}}=\left(\cos{\theta_{i}}, e^{j\chi_{1i}}\cos{\epsilon_{i}}\sin{\theta_{i}}, e^{j\chi_{2i}}\sin{\epsilon_{i}}\sin{\theta_{i}}\right)^{T}, 
\end{equation}
where $j$ is the complex unit. As the weak value is invariant under unitary transformations, we choose to map the pre-selected state to the state $\ket{\psi'_{i}}=\left(1,0,0\right)^{T}$ that is separable in the Majorana representation, $\ket{\Psi'_{i}}=\ket{\phi_i}\ket{\phi_i}$, $\ket{\phi_i}=\ket{\uparrow}$, choosing $\ket{\uparrow}=\left(1,0\right)^T$ and $\ket{\downarrow}=\left(0,1\right)^T$. The unitary operator that maps the pre-selected state to the state $\ket{\psi'_{i}}$ is, 
\begin{eqnarray}
  \label{eq:unitary_first}
\hat{U}^{\left(1\right)} =
  \begin{pmatrix}
	\cos{\theta_{i}} & e^{-j\chi_{1i}}\cos{\epsilon_{i}}\sin{\theta_{i}} & e^{-j\chi_{2i}}\sin{\epsilon_{i}}\sin{\theta_{i}}\\
   \sin{\theta_{i}} & -e^{-j\chi_{1i}}\cos{\epsilon_{i}}\cos{\theta_{i}} & -e^{-j\chi_{2i}}\sin{\epsilon_{i}}\cos{\theta_{i}} \\    
		  0 &-e^{-j\chi_{1i}}\sin{\epsilon_{i}}              & e^{-j\chi_{2i}}\cos{\epsilon_{i}} 
 \end{pmatrix}
 \nonumber\\ 
\end{eqnarray} 
When applying this unitary operator to the system, the post-selected state and the observable are also modified, $\ket{\psi_f}\rightarrow \ket{\psi'_f}$ and $\hat{A}\rightarrow\hat{A}'$. After removing the phase on the first component of the state \footnote{As the expressions of the weak value depend on the projector, $\Pi_{i'}$, this phase has no impact} arising from the application of the observable over the initial state, $\ket{\psi'_{i'}}$ can be written as $\ket{\psi'_{i'}}=\left(\cos{\theta_{i'}},e^{j\chi_{1i'}}\cos{\epsilon_{i'}}\sin{\theta_{i'}}, e^{j\chi_{2i'}}\sin{\epsilon_{i'}}\sin{\theta_{i'}}\right)^{T}$. At this stage, we apply a second unitary operator $\hat{U}^{\left(2\right)}$ that leaves $\ket{\psi'_{i}}$ invariant and takes the state $\ket{\psi'_{i'}}$ to a separable state, 
\begin{eqnarray}
\label{eq:unitary_first}
\hat{U}^{\left(2\right)} =
	\begin{pmatrix}
	1  & 0 & 0\\
	0 & e^{-j\chi_{1i'}}\cos{\alpha} & -e^{-j\chi_{2i'}}\sin{\alpha}\\
  0 &  e^{-j\chi_{1i'}}\sin{\alpha}  & e^{-j\chi_{2i'}}\cos{\alpha}      
	\end{pmatrix},
\end{eqnarray} 
where $\alpha=-\epsilon_{i'}+\arcsin{\left(\tan{\frac{\theta_{i'}}{2}}\right)}$. This unitary transformation maps $\ket{\psi'_{i'}}$ to,
\begin{equation}
\ket{\psi''_{i'}}=\left(\cos{\theta_{i'}}, \sqrt{2\cos{\theta_{i'}}\left(1-\cos{\theta_{i'}}\right)}, 1-\cos{\theta_{i'}}\right)^{T}
\end{equation}
which, in terms of qubits, is $\ket{\Psi''_{i'}}=\ket{\phi_{i'}}\ket{\phi_{i'}}$, with
\begin{equation}
\ket{\phi_{i'}}=\left(\sqrt{\cos{\theta_{i'}}},\sqrt{1-\cos{\theta_{i'}}}\right)^T.
\end{equation}
After applying both unitary transformations ($\hat{U}^{\left(1\right)}$ and $\hat{U}^{\left(2\right)}$), the post-selected state $\ket{\psi''_{f}}$ has the general form, $\ket{\psi''_{f}}=c_0\ket{0}+c_1\ket{1}+c_2\ket{2}$. To obtain the Majorana symmetrized state, one should solve the following polynomial \cite{tamate2011bloch},
\begin{eqnarray}
c_2-\sqrt{2}c_1z+c_0z^2=0
  \label{eq:polynomial_three_level_systems}
\end{eqnarray} 
The polar, $\theta_k$, and azimuthal, $\phi_k$, angles on the Bloch sphere can be calculated from the roots $z_k$ of the polynomial, Eq.~(\ref{eq:polynomial_three_level_systems}). More details on the application of the Majorana symmetric representation can be found in~\ref{section:appendix_2}.\\
Once all the transformations are applied, the three states are easily mapped to the Bloch sphere, 
\begin{eqnarray}
\ket{\psi_i}&\rightarrow&\ket{\Psi_i}=\ket{\uparrow}\ket{\uparrow} \\ \nonumber
\ket{\psi_{i'}}&\rightarrow&\ket{\Psi_{i'}}=\ket{\phi_{i'}}\ket{\phi_{i'}} \\ \nonumber
\ket{\psi_{f}}&\rightarrow&\ket{\Psi_{f}}=\frac{1}{\sqrt{M}}\left(\ket{\phi_{f}^{\left(1\right)}}\ket{\phi_{f}^{\left(2\right)}}+\ket{\phi_{f}^{\left(2\right)}}\ket{\phi_{f}^{\left(1\right)}}\right), 
\end{eqnarray}
with $M=2\left(1+\lvert\bra{\phi_{f}^{\left(1\right)}}\ket{\phi_{f}^{\left(2\right)}}\rvert^2\right)$. The weak value written in terms of the new states is,  
\begin{eqnarray}
A_w&=&\frac{\bra{\psi_{i}}\hat{A}^2\ket{\psi_{i}}}{\bra{\psi_{i}}\hat{A}\ket{\psi_{i}}}\frac{\bra{\phi^{\left(1\right)}_f}\ket{\phi_{i'}}\bra{\phi^{\left(2\right)}_f}\ket{\phi_{i'}}\bra{\phi_{i'}}\ket{\phi_i}^2}{\bra{\phi^{\left(1\right)}_f}\ket{\phi_{i}}\bra{\phi^{\left(2\right)}_f}\ket{\phi_{i}}} \\ \nonumber
&=&\frac{\sqrt{\bra{\psi_{i}}\hat{A}^2\ket{\psi_{i}}}}{\bra{\psi_{i}}\hat{A}\ket{\psi_{i}}}\Pi_{i',w}^{\left(1\right)}\Pi_{i',w}^{\left(2\right)},
\end{eqnarray}
where $\Pi^{\left(n\right)}_{i',w}=\frac{\bra{\phi_f^n}\ket{\phi_{i'}}\bra{\phi_{i'}}\ket{\phi_i}}{\bra{\phi_f^n}\ket{\phi_i}}$. 
The quantity $\bra{\psi_{i}}\hat{A}^2\ket{\psi_{i}}$ is real and positive and $\bra{\psi_{i}}\hat{A}\ket{\psi_{i}}$ is real, therefore, the argument of the weak value is the sum of the arguments of both weak values and an extra phase that is either $0$ or $\pi$, \\
\begin{equation}
\text{arg}A_w=\text{arg}\Pi^{\left(1\right)}_{i',w}+\text{arg}\Pi^{\left(2\right)}_{i',w}-\text{arg}\langle\hat{A}\rangle_{i}. 
\end{equation}
The argument of the weak value of any three-level observable is the sum of two arguments of weak values of projectors of qubits. Each of these arguments is associated to the solid angle on the Bloch sphere of the triangle spanned by the vectors representing the states $\ket{\phi_{i}}$, $\ket{\phi_{i'}}$ and $\ket{\phi_f^n}$,
\begin{equation}
\text{arg}A_w=-\frac{\Omega_{ii'f_1}}{2}-\frac{\Omega_{ii'f_2}}{2}-\text{arg}\langle\hat{A}\rangle_{i}
\end{equation}
\\
The argument of weak values for general observables in three-level systems can be described geometrically in two ways: first, within the complex projective space $\text{CP}^{2}$, which represents the space of three-level state projectors; and second, from four qubits on the Bloch sphere, which corresponds to $\text{CP}^{1}$, the space of two-level state projectors. The geometric phase corresponding to the argument of the weak value for any three-level observable is associated to the symplectic area in $\text{CP}^{2}$ of the triangle spanned by the geodesics connecting the three vectors representing the pre-selected state, the application of the observable over the initial state and the post-selected state. Additionally, this symplectic area can be mapped to a Riemannian area on the Bloch sphere thanks to the Majorana description. In this case, the argument of the weak value is the sum of two arguments that are associated to solid angles on the Bloch sphere.\\ 
Any weak value can be described using a three-level system, only three vectors are involved in the calculations. In consequence, the results presented in this section are pertinent for the weak value of any $N$-level  observable. The argument of the weak value of any $N$-level observable is the sum of the argument of the weak value of two qubit projectors (up to a phase of $0$ or $\pi$). Consequently, in this section we have linked the symplectic area of a triangle in $\text{CP}^{N-1}$ to two solid angles on the Bloch sphere.\\ 
In Fig.~\ref{fig:majorana_and_cp2}, we depict the solid angles linked to the argument of the weak value of a chosen Hermitian operator, which corresponds to the controlled NOT gate, essential to produce entangled states in quantum computing,
\begin{equation}
\label{eq:CNOT}
CNOT=
	\begin{pmatrix}
	1  & 0 & 0 & 0\\
	0 & 1 & 0 & 0\\
  0 & 0 & 0 & 1 \\
	0 & 0 & 1 & 0
	\end{pmatrix}.
\end{equation}
In Fig.~\ref{fig:majorana_and_cp2}a), we represent the three solid angles on the Bloch sphere in the Majorana representation and the three states involved in the weak value. Nevertheless, one can always reduce the size of the system to a three-level system (independently on the initial vector space size). In Fig.~\ref{fig:majorana_and_cp2}b), we depict the two solid angles induced by the argument of the weak value after having reduced the size of the space from $4$ levels (requiring $3$ states) to $3$ levels (involving only $2$ states). \lorena{To reduce the size of the problem, }
\yc{one should follow a Gram-Schmidt orthogonalization procedure from the three states $\ket{\psi_i}$, $\ket{\psi_{i^\prime}}$, and $\ket{\psi_f}$. In the resulting orthogonal basis, the three states have at most 1, 2, and 3 non-zero components, respectively. Then, we can completely describe these three states in the Majorana representation restricted to the three-dimensional subspace generated by the three first vectors of the basis. The required unitary operator performs thus the basis change from the original basis to the new one constructed.} In Fig.~\ref{fig:majorana_and_cp2}c), the geodesic triangle between the three involved vectors is represented in the complex projective space $\text{CP}^2$. To do so, we use the spherical octant projection. Each point of the octant is associated to a torus formed by the two phase components of the Hilbert space, $\chi_1$ and $\chi_2$, $\ket{\psi}=\left(|\psi_0|, e^{j\chi_1}|\psi_1|,  e^{j\chi_2}|\psi_2|\right)^T$ \cite{mallesh1997generalized}. The state $\ket{\psi}$ is projected to the real point $\vec{q}=\left(|\psi_1|, |\psi_2|, |\psi_0|\right)$.  We also depicted the geodesics between each pair of states in three dimensions. The symplectic area is a contour integral along these geodesics, but it cannot be directly represented. The geodesics in $\text{CP}^2$ do not correspond to great circles on the sphere $S^7$. Moreover, after applying the Majorana representation to the geodesic, it does not correspond to great circles on the Bloch sphere. Between separable states, geodesics correspond to circular segments on the Bloch sphere \cite{mittal2022geometric}. However, they can have very complicated shapes when it comes to not separable states.\\
\section{Weak values of three-level systems: Spin-1}\label{section:spin_1}
\begin{figure} [h!]
\centering 
\includegraphics[width=0.5\textwidth]{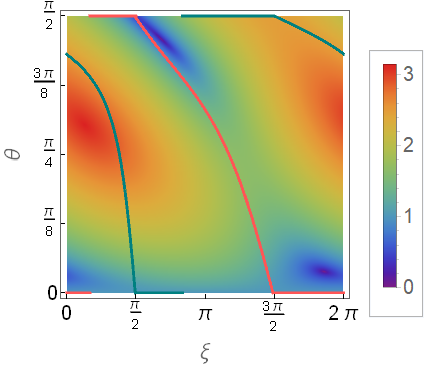}
 \caption{\lorena{Color map of of the angle between the vectors representing the post-selected state, $\Vec{f_1}$ and $\Vec{f_2}$, on the Bloch sphere as a function of $\theta$ and $\xi$. The red line indicates the angle $\theta_{\text{max}}\left(\xi\right)$, where the modulus of the weak value is maximized for fixed values of $\xi$. The green line represents the angle $\theta_{\text{min}}\left(\xi\right)$, where the modulus of the weak value is minimized for fixed values of $\xi$.} \label{fig:entropy}}
\end{figure}

The spin, describing the intrinsic angular momentum of particles, has a central role in quantum physics. The spin operator depends on the type of particle. The Pauli matrices, the chosen generators of $\text{SU}\left(2\right)$, describe the spin-1/2 \cite{eisberg1985quantum}. In the case of spin-1, the operators can be described in terms of generators of $\text{SU}\left(3\right)$. The spin operators along the three different axes are detailed in terms of the Gell-Mann matrices as, $\hat{S}_x=\frac{1}{\sqrt{2}}\left(\hat{\lambda}_1+\hat{\lambda}_6\right)$, $\hat{S}_y=\frac{1}{\sqrt{2}}\left(\hat{\lambda}_2+\hat{\lambda}_7\right)$, $\hat{S}_z=\frac{1}{2}\left(\hat{\lambda}_3+\sqrt{3}\hat{\lambda}_8\right)$ \cite{binicioglu2007entanglement}, where the Gell-Mann matrices are defined in~\ref{section:appendix_1} and $\hbar=1$.\\
In several experiments, the weak value of the spin operators has a central role \cite{aharonov1988result, romito2008weak}. The real and imaginary parts of the weak values of spin-$1/2$ operators have been theoretically studied, along with their modulus and argument \cite{aharonov1988result, duck1989sense, cormann2016revealing, sjoqvist2006geometric}. As the spin direction can be represented directly on the Bloch sphere, the situation is easy to visualize. However, the weak values of the spin-1 operators were much less studied, specially from a geometrical point of view. One possible method is its study in terms of vectors in the complex projective space $\text{CP}^2$, with a generalization of the Bloch sphere \cite{ballesteros2022geometrical}. Here, we focus on the description of weak value of the spin-1 operator on the Bloch sphere using the Majorana formalism introduced in the previous sections. \\
Let us consider the weak value of a linear combination of the three components of the spin, $\vec{\hat{S}}=n_x\hat{S}_x+n_y\hat{S}_y+n_z\hat{S}_z$. Without loss of generality, by setting an appropriate reference point, we rotate the direction $\vec{n}=\left(n_x, n_y, n_z\right)^T$  into $\vec{n}=\left(0,0,1\right)$. In consequence, we focus on the study of the weak value of $\hat{S}_z$, ${S_z}_w$. The general pre- and post-selected states have $4$ independent parameters each. To simplify the studied case, the pre-selected state is chosen to be,
\begin{equation}
\ket{\psi_i}=\frac{1}{\sqrt{6}}\left(2,1,j\right)^T,
\end{equation}
where we set the parameters in Eq.~(\ref{eq:expression_psi_i_in_terms_of_vectors}) to $\epsilon_i=\frac{\pi}{4}$, $\chi_{1i}=0$, $\chi_{2i}=\frac{3\pi}{2}$, and $\theta=\arccos{\sqrt{\frac{2}{3}}}$, a simple state, but not a trivial one. In the case of the post-selected state, only two parameters are fixed, $\epsilon_f=\frac{\pi}{4}$, and $\chi_{2f}=0$, 
\begin{equation}
\ket{\psi_f}=\left(\cos{\theta}, \frac{1}{\sqrt{2}}\sin{\theta}e^{j\xi}, \frac{1}{\sqrt{2}}\sin{\theta}\right)^T. 
\end{equation}
These states provide a system with two independent parameters, $\theta$ and $\xi$. The application of the spin operator to the pre-selected state is 
\begin{equation}
\ket{\psi_{i'}}=\frac{1}{\sqrt{5}}\left(2,0,-j\right).
\end{equation}
Applying the appropriate unitary operators, the initial state is moved to $\ket{\psi''_i}=\left(1,0,0\right)\rightarrow\ket{\Psi_i}=\ket{\uparrow}\ket{\uparrow}$ and the state $\ket{\psi_{i'}}$ to 
\begin{equation}
\ket{\psi''_{i'}}=\left(\sqrt{\frac{3}{10}}, \sqrt{-\frac{3}{5}+\sqrt{\frac{6}{5}}}, 1-\sqrt{\frac{3}{10}}\right)^T \rightarrow \ket{\Psi_{i'}}=\ket{\phi_{i'}}\ket{\phi_{i'}}, 
\end{equation}
where 
\begin{equation}
\ket{\phi_{i'}}=\left(\left(\frac{3}{10}\right)^{\frac{1}{4}},\sqrt{1-\sqrt{\frac{3}{10}}}\right)^T.
\end{equation}
Making use of these states and applying the Majorana representation to the post-selected state (in Eq.~(\ref{eq:general_polynomial}) one finds the Majorana polynomial that should be solved), we study the argument of the weak value of the spin-1 as the sum of two arguments of two-level projectors that are associated to two solid angles on the Bloch sphere. \\
\begin{figure} [h!]
\centering 
\includegraphics[width=0.5\textwidth]{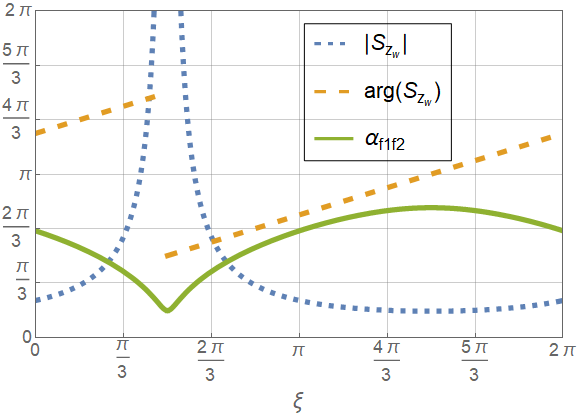}
  \caption{\lorena{Representation of the maximum}\yc{ value of the}\lorena{ modulus of the weak value as a function of $\xi$ (blue), the argument of the weak value at $\theta_{\text{max}}\left(\xi\right)$ where the modulus is maximized (green), and the angle $\alpha_{f1f2}$ between $\vec{f}_1$ and $\vec{f}_2$ at $\theta_{\text{max}}\left(\xi\right)$ as a function of $\xi$ (orange)}	
\label{fig:max_abs_angle_f1_f2_spin_1}}
\end{figure}
One of the most useful characteristics of weak values is their ability to amplify minute phenomena thanks to  their unbounded property. Identifying the behavior of the argument of the weak value when the absolute value tends to infinity is essential due to both the discontinuities that appear in that range and their usefulness. \\
In Fig.~\ref{fig:max_abs_angle_f1_f2_spin_1}, we represent the maximum value of the modulus of the weak value of $\hat{S}_z$ for each value of $\xi$. The maximum of the modulus of the weak values takes place for a determined $\theta_{\text{max}}\left(\xi\right)$. We use this value to plot the argument of the weak value at $\theta_{\text{max}}\left(\xi\right)$ in terms of $\xi$. We also depict the angle between the two stars on the Bloch sphere representing the post-selected state, $\Vec{f}_1$ and $\Vec{f}_2$ at $\theta_{\text{max}}\left(\xi\right)$ as a function of $\xi$. The angle between the two vectors on the Bloch sphere represents an entanglement measurement of the two-qubit state. If the angle between the vectors is $0^{\circ}$, the state is separable and thus the entropy of entanglement is $0$. On the opposite side, if the angle between the two vectors is $180^{\circ}$, the state is a maximally entangled Bell state. The modulus of the weak value presents a vertical asymptote at $\xi=\frac{\pi}{2}$ because the initial and final states are then orthogonal. At the divergence point, the argument of the weak value presents a $\pi$ jump. This behavior is typical of the argument of the weak value when there is a divergence in the modulus \cite{cormann2017geometric}.\\ 
The two vectors on the Bloch sphere associated to the final state, $\vec{f}_1$ and $\vec{f}_2$, are the closest, $29.42^{\circ}$, where the maximum value of the modulus tends to infinity.  Both the initial state and the application of the operator over the initial state present an entropy of entanglement equal to $0$, as the states are separable. Hence, the angle between $\vec{f}_1$ and $\vec{f}_2$ represents the total entanglement of the system. \\
Having a minimum of entropy of entanglement at the divergence in the modulus of the weak values is counter-intuitive at first. Anomalous weak values are a proof of contextuality \cite{pusey2014anomalous}, a characteristic of non-classicality. Therefore, it could have been expected to find a maximum in the entanglement, which is also a characteristic of non-classicality, at the most anomalous weak value (divergence). To clarify if this is an intrinsic characteristic of the system, we depict the value of the angle between the vectors $\Vec{f_1}$ and $\Vec{f_2}$ for all values of $\theta$ and $\xi$ in Fig.~\ref{fig:entropy}. We also include the value of $\theta$ at the maximum of the modulus of the weak value, $\theta_{\text{max}}\left(\xi\right)$ (red line). We plot the same line for the minimum of the weak value $\theta_{\text{min}}\left(\xi\right)$ (green line). There are two absolute minima of the entropy of entanglement. None of them is at the maximum of the modulus of the weak value (red line in the plot). However, the maximum of modulus of the weak value is always located close to the minimum of the entanglement, as it follows the bottom of the valley of minimal entanglement on Fig.~\ref{fig:entropy} (slightly to the left). A very similar correlation links the minimum of the modulus of the weak value (green line) and the maximum of the entanglement. The trends are very similar, but slightly shifted. This behavior is very intriguing due to the correlation of the anomalous weak values and non-classicality. We think this should be further explored in the future.\\     
Using the Majorana representation, we studied different aspects of the weak values, such as the entropy of entanglement. We noticed that an interesting behavior occurs: a maximum (minimum) of the entanglement is near a minimum (maximum) of the modulus of the weak value. Only the Majorana approach allows this analysis. The entanglement of the Majorana stars has a clear meaning when the initial state is an actual 2-particle system in a symmetric state. Otherwise, the results can be interpreted following a single-particle entanglement formalism \cite{van2005single}. Further details on the analysis of the spin-1 can be found in~\ref{section:appendix_3}
\section{Conclusions}\label{section:conclusions}
We applied the Majorana symmetric representation to study the geometry of the argument of weak values of $N$-level general observables on the Bloch sphere. The argument of the weak value can provide significant information in the study of interferometric phenomena with post-selection. Moreover, it is connected to the ratio 
between the optimal conditional estimate of the observable and the inaccuracy of the estimate. The weak value of any observable is proportional to the weak value of an effective projector that is defined as the normalized application of the observable over the pre-selected state. The constant of proportionality is real. Hence, the argument of the weak value of any observable is the argument of the weak value of a projector modulo $\pi$. \\
The modulus of the weak value of a general observable is the product of $N-1$ moduli of weak values of projectors in the complex projective space $\text{CP}^{1}$ that describe qubits, and of constants that are independent on the post-selected state. The argument of the weak value of any observable is the sum of $N-1$ arguments of weak values of projectors in $2$-level systems, plus a phase that is either $0$ or $\pi$. Each of these arguments represents a solid angle on the Bloch sphere. This description generalizes to the argument of any third-order Bargmann invariant and of the Kirkwood-Dirac quasi-probabilty distribution.\\
Any weak value depends only on three states. Thus, applying different unitary operators, it is possible to map these states to a three-level system, giving a special importance to the qutrit case. Doing so, we map a symplectic area in the complex projective space $\text{CP}^{N-1}$ that represents the $N$-dimensional quantum state manifold to a sum of two solid angles, instead of $N-1$, on the Bloch sphere (up to a constant that is either $0$ or $\pi$). The solid angles on the Bloch sphere are determined by the great circles between the four qubit vectors (the two degenerate states associated to the initial state, the two degenerate states linked to the observable, and the two entangled states describing the final state in the symmetric Majorana representation). However, these great circles are not representations of geodesics between the states in the complex projective space $\text{CP}^{N-1}$. This is an advantage since great circle arcs are easy to determine, contrary to the Majorana representation of the true geodesics of the quantum state manifold.\\
We applied these results to the spin-1 operator for anomalous weak values in the region of weak value amplification. Using a specific case, we studied the argument of the weak value when the modulus tends to infinity (asymptotic behavior). We found that when the weak value diverges, the angle between the two vectors representing the post-selected state on the Bloch sphere presents a constrained minimum. The angle between the two qubits representing the post-selected state on the Bloch sphere gives a measure of the total entanglement of the system. The maximum value of the modulus is for any value of the angle $\xi$ near the minimum of entanglement. We should highlight that only the study in terms of Majorana stars makes possible an entanglement discussion. The physical meaning of entanglement when the initial state is a 2-particle system in a symmetric state is straightforward. However, when this is not the case, studies should be performed in the light of single-particle entanglement \cite{van2005single}. 

\section*{Acknowledgments}
Y.C. is a Research Associate of the Fund for Scientific Research F.R.S.-FNRS. This research was supported by the Action de Recherche Concertée WeaM at the University of Namur (19/23-001).
\appendix

\section{Symmetric Majorana representation}\label{section:appendix_2}
In this paper, we used the symmetric Majorana representation of the complex projective space $\text{CP}^N$, which should not be confused with the Majorana representation of spinors. This representation maps $N$-level quantum states to $N-1$ stars on the Bloch sphere. To do so, it associates the basis of $N$-level systems with the symmetric tensorial products of two-level states \cite{devi2012majorana}.\\   
Let us consider a four-level system. The orthonormal basis, which spans all possible states, consists of the states $\ket{0}$, $\ket{1}$, $\ket{2}$, and $\ket{3}$. The Majorana representation maps the four-level states to the symmetric three-qubit states as follows:
\begin{eqnarray}\label{eq:app_Majo_1}
\ket{0}&\rightarrow& \ket{\Psi_0}=\ket{\uparrow}\ket{\uparrow}\ket{\uparrow} \\ \nonumber
\ket{1}&\rightarrow& \ket{\Psi_1}=\frac{1}{\sqrt{3}}\left(\ket{\downarrow}\ket{\uparrow}\ket{\uparrow}+\ket{\uparrow}\ket{\downarrow}\ket{\uparrow}+\ket{\uparrow}\ket{\uparrow}\ket{\downarrow}\right)\\ \nonumber
\ket{2}&\rightarrow& \ket{\Psi_2}=\frac{1}{\sqrt{3}}\left(\ket{\downarrow}\ket{\downarrow}\ket{\uparrow}+\ket{\downarrow}\ket{\uparrow}\ket{\downarrow}+\ket{\uparrow}\ket{\downarrow}\ket{\downarrow}\right)\\ \nonumber
\ket{3}&\rightarrow& \ket{\Psi_3}=\ket{\downarrow}\ket{\downarrow}\ket{\downarrow}, \\ \nonumber
\end{eqnarray}\\
where $\ket{\uparrow}$ and $\ket{\downarrow}$ represent the spin up and spin down states of the qubit, i. e. the eigenstates of the $\hat{\sigma}_z$ Pauli operator associated to the eigenvalues $+1$ and $-1$, respectively. \yc{The four states $\ket{\Psi_i}$ are symmetric under permutation of the single qubit spaces.} These results can be found by solving the polynomial given in Eq.~(\ref{eq:general_polynomial}). The order of the basis can be chosen arbitrarily. Any four-level state can be transformed into \yc{a symmetric state of three qubits} by forming a linear combination of the states $\ket{\Psi_0}$, $\ket{\Psi_1}$, $\ket{\Psi_2}$, and $\ket{\Psi_3}$. Alternatively, the parameters defining the three-qubit state can be determined using the polynomial given in Eq.~(\ref{eq:general_polynomial}). \\ 
The mapping between an $N$-level state and symmetric $(N-1)$-qubit states can be performed as follows:
\begin{eqnarray}\label{eq:app_Majo_2}
\ket{0}&\rightarrow& \ket{\Psi_0}=\underbrace{\ket{\uparrow}\ket{\uparrow}....\ket{\uparrow}}_{N-1} \\ \nonumber
\ket{1}&\rightarrow& \ket{\Psi_1}=\frac{1}{\sqrt{N-1}}\sum_{P}\hat{P}\ket{\downarrow}\underbrace{\ket{\uparrow}\ket{\uparrow}...\ket{\uparrow}}_{N-2}\\ \nonumber
\vdots \\ \nonumber
\ket{N-2}&\rightarrow& \ket{\Psi_{N-2}}=\frac{1}{\sqrt{N-1}}\sum_{P}\hat{P}\ket{\uparrow}\underbrace{\ket{\downarrow}\ket{\downarrow}...\ket{\downarrow}}_{N-2}, \\ \nonumber
\ket{N-1}&\rightarrow& \ket{\Psi_{N-1}}=\underbrace{\ket{\downarrow}\ket{\downarrow}...\ket{\downarrow}}_{N-1}, \\ \nonumber
\end{eqnarray}\\
where $P$ runs through all the permutations $\hat{P}$ of the states $\ket{\uparrow}$ and $\ket{\downarrow}$. In general, it is possible to calculate the Majorana stars of any general $N$-level state using the polynomial described in Eq.~(\ref{eq:general_polynomial}). The states $\ket{\Psi_k}$ in (\ref{eq:app_Majo_1}) and (\ref{eq:app_Majo_2}) are also know as the Dicke completely symmetric states $\mathcal{D}_k^{n}$, with $n=N-1$ the number of qubits involved and $k$, an index running from 0 to $n$, as defined in (\ref{eq.Dicke}).

\section{Gell-Mann matrices}\label{section:appendix_1}
The order of the Gell-Mann matrices used in this paper is\\
\begin{eqnarray}
\hat{\lambda}_1=
\begin{pmatrix}
0 & 1 & 0 \\
1 & 0 & 0 \\
0 & 0 & 0
\end{pmatrix}
\hspace{1 cm}
\hat{\lambda}_2=
\begin{pmatrix}
0 & -i & 0 \\
i & 0 & 0 \\
0 & 0 & 0
\end{pmatrix}
\\ \nonumber
\hat{\lambda}_3=
\begin{pmatrix}
1 & 0 & 0 \\
0 & -1 & 0 \\
0 & 0 & 0
\end{pmatrix}
\hspace{1 cm}
\hat{\lambda}_4=
\begin{pmatrix}
0 & 0 & 1 \\
0 & 0 & 0 \\
1 & 0 & 0
\end{pmatrix}
\\ \nonumber
\hat{\lambda}_5=
\begin{pmatrix}
0 & 0 & -i \\
0 & 0 & 0 \\
i & 0 & 0
\end{pmatrix}
\hspace{1 cm}
\hat{\lambda}_6=
\begin{pmatrix}
0 & 0 & 0 \\
0 & 0 & 1 \\
0 & 1 & 0
\end{pmatrix}
\\ \nonumber
\hat{\lambda}_7=
\begin{pmatrix}
0 & 0 & 0 \\
0 & 0 & -i \\
0 & i & 0
\end{pmatrix}
\hspace{1 cm}
\hat{\lambda}_8=\frac{1}{\sqrt{3}}
\begin{pmatrix}
1 & 0 & 0 \\
0 & 1 & 0 \\
0 & 0 & -2
\end{pmatrix}.
\end{eqnarray}

\section{Advanced analysis of spin-1 systems}\label{section:appendix_3}
\begin{figure} [h!]
\centering 
\includegraphics[width=0.5\textwidth]{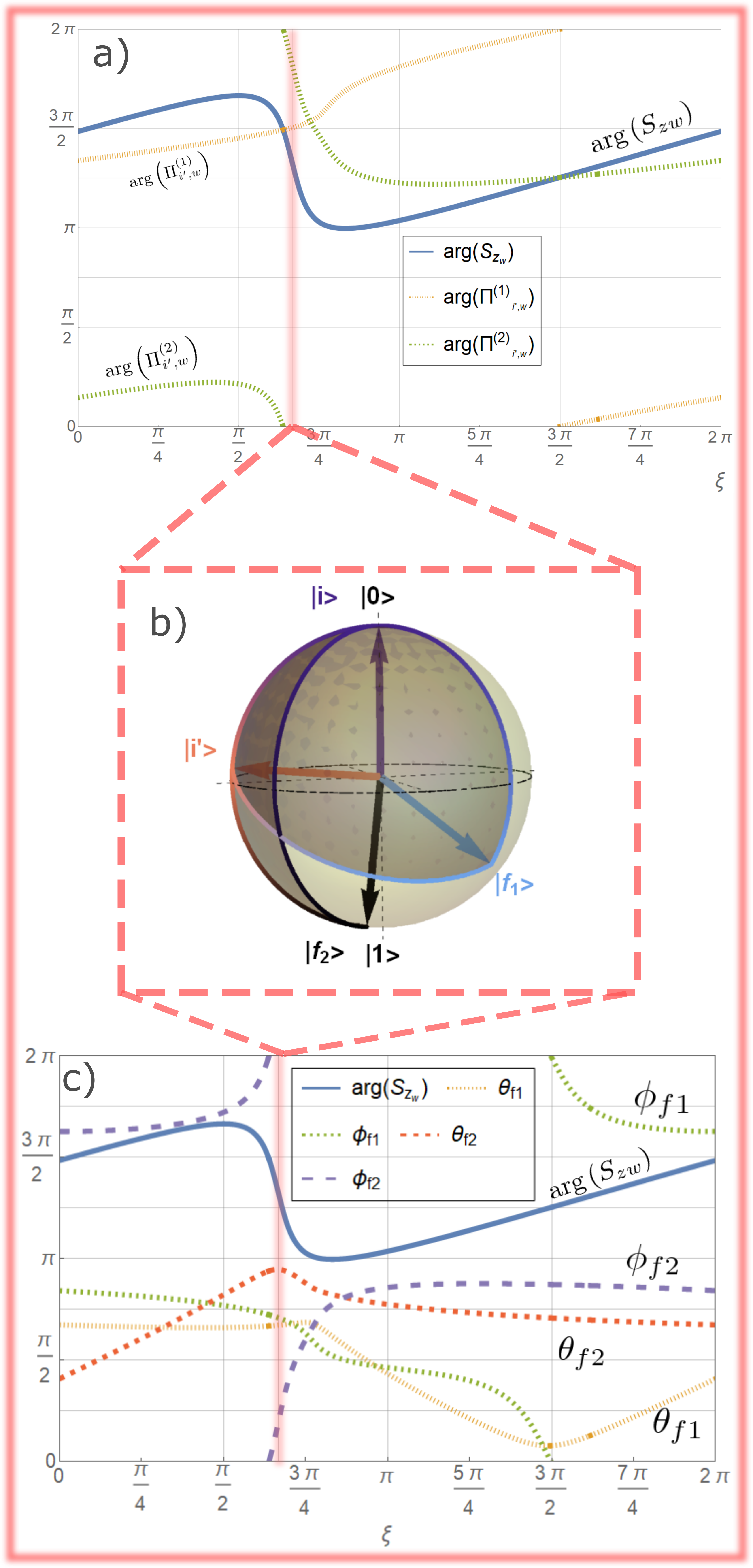}
\caption{a) Argument of the weak value of the spin operator, ${S_z}_w$ \lorena{as a function} of $\xi$, \lorena{along with} \lorena{the arguments of the weak values} ${\Pi^{\left(1\right)}_{i',w}}$ and ${\Pi^{\left(2\right)}_{i',w}}$ \lorena{as a function of} $\xi$ for $\theta=\frac{\pi}{2}-0.2$. b) Solid angles $\Omega_{ii'f_1}$ and $\Omega_{ii'f_2}$ \lorena{on the Bloch sphere} for $\theta=\frac{\pi}{2}-0.2$ and $\xi=2.09$. c) Argument of the weak value of $\hat{S}_z$ and polar \lorena{($\theta$)} and azimuthal \lorena{($\phi$)} angles of the vectors representing the post-selected state on the Bloch sphere for $\theta=\frac{\pi}{2}-0.2$, \lorena{as a function of} $\xi$. A vertical line \lorena{is included} in a) and c) \lorena{at $\xi=2.09$}, where the modulus of the weak value is \lorena{maximized}. \label{fig:behaviour_pi_3_002}}
\end{figure}

\begin{figure} [h!]
\centering 
\includegraphics[width=0.5\textwidth]{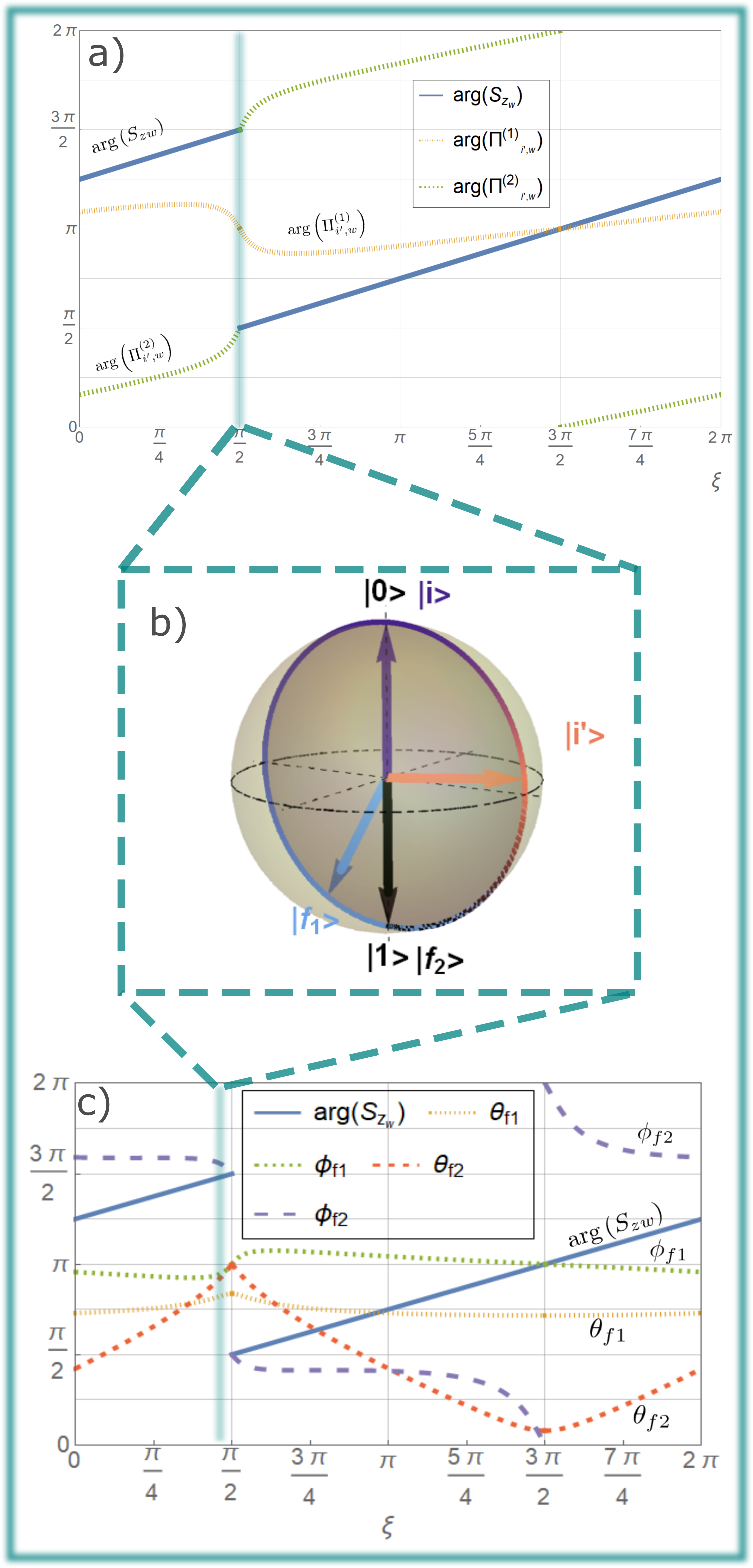}
  \caption{a) Argument of the weak value of of the spin operator, ${S_z}_w$ \lorena{as a function} of $\xi$, \lorena{along with} \lorena{the arguments of the weak values} ${\Pi^{\left(1\right)}_{i',w}}$ and ${\Pi^{\left(2\right)}_{i',w}}$ \lorena{as a function of} $\xi$ \lorena{for} $\theta=\frac{\pi}{2}-10^{-11}$. b) Solid angles  $\Omega_{ii'f_1}$ and $\Omega_{ii'f_2}$ \lorena{on the Bloch sphere} for $\xi=\frac{\pi}{2}$ and $\theta=\frac{\pi}{2}$. c) Argument of the weak value of $\hat{S}_z$ and polar \lorena{($\theta$)} and azimuthal \lorena{($\phi$)} angles of the vectors representing the post-selected state on the Bloch sphere for $\theta=\frac{\pi}{2}$,  \lorena{as a function of} $\xi$. A vertical line \lorena{is included} in a) and c) at \lorena{at $\xi=\frac{\pi}{2}$}, where the modulus of the weak value is \lorena{maximized}.\label{fig:behaviour_pi_2}}
\end{figure}

The Majorana representation is very useful to visualise the geometric phase in an intuitive way, as a function of an appropriate dynamical parameter (such as a variable determining the pre- or post-selected states in a weak measurement). We depict an example in Fig.~\ref{fig:behaviour_pi_2} and Fig.~\ref{fig:behaviour_pi_3_002}. We note that parts a) and c), which essentially provide the spherical coordinates of the Majorana stars in relationship to the weak value, may not appear particularly easy to follow as static graphs. However, parts b) illustrate clearly the geometric configuration for a particular value of the parameter. Thus, in practice, the dynamical evolution of the stars (i.e. their correlated trajectories on the Bloch sphere) and the associated geometric phase as a function of the relevant parameter would be neatly described on a computer screen, facilitating substantially the analysis (but the evolution cannot be represented dynamically in this printed paper). Let us analyse more precisely these figures, depicting further the case furnished in the section studying the spin-1 application, section~\ref{section:spin_1}.\\
In Fig.~\ref{fig:behaviour_pi_3_002} and Fig.~\ref{fig:behaviour_pi_2}, we depict the evolution of the argument of the weak value in the Majorana representation in terms of $\xi$ (a), the representation on the Bloch sphere of the solid angles associated to the argument of the weak value for the maximum of the modulus of the weak value (b), and the evolution of the angles on the Bloch sphere as a function of $\xi$ (c). In Fig.~\ref{fig:behaviour_pi_3_002}, we represent a case with $\theta$ smaller than at the divergence ($\theta=\frac{\pi}{2}-0.2$) and in Fig.~\ref{fig:behaviour_pi_2}, a case very near the asymptote, ($\theta=\frac{\pi}{2}-10^{-11}$). \\
In Fig.~\ref{fig:behaviour_pi_3_002}a), one can perceive that, around $\theta=\frac{\pi}{2}$, the slope of the function is quite big. The closer the $\theta$ is from the divergence case, the larger the slope. In Fig.~\ref{fig:behaviour_pi_2}a), we represent the extreme case, when $\theta=\frac{\pi}{2}$. There, the slope is infinite, as the argument presents a $\pi$ jump. In the first case, Fig.~\ref{fig:behaviour_pi_3_002}, the big slope in the argument of the weak value occurs when $\text{arg}\Pi^{\left(2\right)}_{i',w}$ passes by $0$, so that there is no discontinuity. In the second case, Fig.~\ref{fig:behaviour_pi_2}, the argument varies linearly with $\xi$, except at the point of the maximum of the modulus of the weak value, where it exhibits a $\pi$ jump. To observe a smooth movement of the star on the Bloch sphere (without change of sense of the movement), a $\pi$ jump should be present in the function of azimuthal angle, $\phi$.
\\
In figures b) of Fig.~\ref{fig:behaviour_pi_3_002} and Fig.~\ref{fig:behaviour_pi_2}, we represent the solid angles on the Bloch sphere associated to the argument of the weak value. In each figure, the solid angles correspond to the case in which the modulus of the weak value is maximum, $\{\theta=\pi/2-0.2, \xi=2.09\}$ and $\{\theta=\frac{\pi}{2}, \xi=\frac{\pi}{2}\}$ respectively. The value of the maximum is highlighted with a vertical line (pink in Fig.~\ref{fig:behaviour_pi_3_002} and green in Fig.~\ref{fig:behaviour_pi_2}). Far from the divergence, Fig.~\ref{fig:behaviour_pi_3_002}, there are clearly two solid angles. However, very close to the divergence, Fig.~\ref{fig:behaviour_pi_2},  all the vectors are nearly on the same plane on the Bloch sphere. One of the qubit states representing the post-selected state in the Majorana representation is orthogonal to the qubit state representing the initial state, a condition required for the appearance of a divergence. When the initial and final states are orthogonal, the great circle between $\Vec{i}$ and $\Vec{f}_2$ is not unique as there are different paths with the same distance. \\
In Fig.~\ref{fig:behaviour_pi_3_002}c,~\ref{fig:behaviour_pi_2}c, we depict both the azimuthal and the polar angles of the two qubits representing the final state, at the divergence position, $\theta=\frac{\pi}{2}$ in Fig.~\ref{fig:behaviour_pi_2} and at a smaller value of $\theta$, Fig.~\ref{fig:behaviour_pi_3_002}. In Fig.~\ref{fig:behaviour_pi_3_002}c, the polar angle $\theta_1$ is approximately constant from $\xi=0$ until the maximum of the modulus of the weak value (vertical line), $\xi=2.09$. The maximum occurs at a value a bit larger than $\xi=2.09$. Then, it decreases, presenting a minimum at $\xi=\frac{3\pi}{2}$, where the azimuthal angle, $\phi_1$, passes by $0$. After that point, the polar angle increases until reaching the initial value. The polar angle of one of the qubits, $\theta_2$, representing the post-selected state has a maximum at the position of the maximum of the modulus of the weak value (vertical line), where it is orthogonal to the pre-selected state, $\bra{f_2}\ket{i}=0$. The polar angle of the other qubit, $\theta_1$, is almost constant in terms of $\xi$ for $\theta=\frac{\pi}{2}$. It exhibits a smooth maximum at $\xi=\frac{\pi}{2}$. At this point the two polar angles are the closest. In Fig.~\ref{fig:behaviour_pi_2}c, $\phi_2$ has a $\pi$ jump at $\xi=\frac{\pi}{2}$, where the divergence takes place.\\
\\

\section*{References}
\bibliographystyle{iopart-num-YC}
\bibliography{Journal_AbbreviationDB.bib, biblio}

\end{document}